\def\Id{{\rm 1\kern-.3em I}}

\documentclass[%
    ,final            
    ,a4paper          
  ]
  {aipproc}

\layoutstyle{6x9}

\begin{document}

\title{
	Quark model description of hadrons
}

\author{Bernard Metsch}{
  address={
	Helmholtz-Institut f\"ur Strahlen-- und Kernphysik\\
	Rheinische Friedrich--Wilhelms--Universit\"at Bonn\\
	Nu{\ss}allee 14-16, D-53115 Bonn, Germany\\
	E-mail: \texttt{metsch@itkp.uni-bonn.de}
	}
}

\begin{abstract}
In this contribution I will try to give an overview of what has been
achieved in constituent quark models of mesons and baryons by a
comparison of some selected results from various \textit{ans\"atze}
with experimental data. In particular I will address the role of
relativistic covariance, the nature of the effective quark forces, the
status of results on electromagnetic and strong-decay observables
beyond the mere mass spectra, as well as some unresolved issues in
hadron spectroscopy.
\end{abstract}

\maketitle


\section{Introduction}

Although some appreciable progress has been made in \textit{ab initio}
calculations of low-lying baryon resonances within the lattice gauge
approach, still the only comprehensive description of the complete
known spectrum of hadrons (focussing on light quark flavours) with
masses up to 3 GeV, which addresses such issues as linear
Regge-trajectories, parity doublets in the baryon spectrum, the
conspicuous structure of scalar excitations of hadrons, is in fact the
constituent quark model, which assumes that the majority of meson and
baryon excitations can be effectively described as $q\bar{q}$-- and
$q^3$-- bound states of (constituent) quarks and that the coupling to
more complicated configurations (such as strong decay channels) can be
treated perturbatively. Although recent experimental findings hint at
the existence of exotic meson and baryon resonances, this scheme at
least constitutes a framework to judge what is to be considered as
exotic.

Since quarks, even when adopting constituent, effective quark masses, 
move in hadrons with velocities which are a significant fraction of the
velocity of light and most non-static observables involve processes at
rather large momentum transfers, the quark model description should be
based on the usual concepts of quantum field theories. In spite of
this, traditionally the quark dynamics in quantitative constituent 
quark models has been formulated on the basis of the non-relativistic 
Schroedinger equation and relativistic corrections have at best been 
parameterized. Recent calculations on electromagnetic form factors 
elucidated the role of Poincare invariance in calculating 
electromagnetic currents.  

The ultimate goal of any hadron model is to obtain a unified
description of
\begin{itemize}
\item
Mass spectra of (\textit{e.g.} light-flavoured) hadrons from the
ground states up to the highest masses < 3 GeV and highest angular
momenta $J<8$ observed, addressing such isuues as: Regge-trajectories,
scalar excitations, (pseudo)scalar mixings (for mesons), parity
doublets (for baryons), undetected resonances, etc.
\item
electroweak properties, such as electroweak form factors, radiative
decays and transitions, semi-leptonic weak decays, etc.
\item
strong (two-body) decays and interactions.
\end{itemize}
Even within the framework of the constituent quark model, the various
approaches found in the literature do not only differ appreciably with
respect to there scope, but also in the modelling of the effective
quark interactions used and in the assumptions concerning the
dynamical equations. Here we can distinguish between (a) field
theoretical approaches, which implement relativistic covariance in the
basic set-up, such as: Lattice-gauge theory of QCD,
Dyson-Schwinger/Bethe-Salpeter approaches relying on a parametrization
of the infrared gluon propagator, see \textit{e.g.}~\cite{mar03},
instantaneous approximations to this, on the basis of a
parametrization of confinement and using instanton-induced
interactions, which allows for addressing the complete light-flavoured
hadron spectrum and not merely the ground and some lower excited
states and (b) quantum mechanical approaches on the basis of the
Schr\"odinger equation with relativistic corrections using confinement
potentials and effective quark interactions based (alternatively) on
O(ne) G(luon) E(xchange), see \textit{e.g.}~\cite{cap00} or
G(oldstone) B(oson) E(xhange), see \textit{e.g.}~\cite{ple03}. Here
Dirac's instant--, point-- or front-- formulation of relativistic
quantum mechanics is invoked to subsequently calculate various
currents.

\section{Mesons}

In the following we will sketch the various assumptions and
approximations made in constituent quark models by focussing on
mesons: Adopting the
framework of quantum field theory mesons are described as bound 
$q\bar{q}$ states with $M^2=\bar{P}^2$, 
described by the Bethe-Salpeter amplitude
$\chi_{\alpha\beta}(x_1,x_2) 
:= 
\langle 0 | T\left[\psi_\alpha(x_1)\bar{\psi}_\beta(x_2)\right]
| \bar{P} \rangle$\,, 
which enters in a set of coupled equations which mutually determine
the full propagators for the fermions and exchange bosons and the
dressed vertex functions involved. In practise one truncates this set
of equations by making an \textit{Ansatz} for some $n$-point function
and solving the equations (Bethe-Salpeter-Equation (BSE) for two
particles or the Dyson-Schwinger-equation (DSE) for the self-energy)
of lower order. In particular, based on an effective gluon propagator
with a specific infrared behaviour this leads to the
renormalization-group-improved rainbow-ladder approach~\cite{mar03},
of which we will quote some interesting results. In a simplified
\textit{Ansatz} one can refrain from solving the DSE and assume that
the fermion propagator has the free form
$S(p) \approx i\left[\gamma^\mu p_\mu-m+i\varepsilon\right]^{-1}$ 
and
to account for the self-energy contributions by introducing a
constituent mass $m$. Furthermore one could assume that the
irreducible interaction kernel is given by a single gluon exchange
(OGE) in Coulomb gauge, possibly with a running coupling, 
where a Coulomb part of the interaction is
instantaneous, and thus in the no-retardation limit $k^2\to -|\vec k|^2$
arrive at an instantaneous OGE--potential. Such instantaneous
interaction kernels allow for a parametrization of confinement by a
string-like potential and, defining the Salpeter-Amplitude as 
$\Phi(\vec{p}) =
\left.\int\!\!\frac{dp^0}{2\pi}\chi(p^0,\vec{p}\right|_{(P=M,\vec 0)}$\,,
one then arrives at the Salpeter-Equation (instantaneous Bethe-Salpeter
Equation)
 \begin{eqnarray}\label{eqn:Sal}
\Phi(\vec p) & = & \int \frac{d^3 p}{(2\pi)^3} \frac{
        \Lambda_1^-(\vec p) \gamma^0 [V(\vec p,\vec p')\Phi(\vec p')]
        \gamma^0 \Lambda_2^+(-\vec p)}{M+\omega_1+\omega_2}
\nonumber \\
        & - & \int \frac{d^3 p}{(2\pi)^3} \frac{
        \Lambda_1^+(\vec p) \gamma^0 [V(\vec p,\vec p')\Phi(\vec p')]
        \gamma^0 \Lambda_2^-(-\vec p)}{M-\omega_1-\omega_2}\,,
\end{eqnarray}
with the projectors $\Lambda_i^\pm(\vec p)=(\omega_i(\vec p)\pm
H_i(\vec p))/2\omega_i(\vec p)$, the Dirac Hamiltonian $H_i(\vec
p)=\gamma^0(\vec \gamma \cdot \vec p + m_i)$ and where $\omega_i(\vec
p) = \sqrt{m_i^2+\vec p^2}$. If one now drops the first term on the
r.h.s. of Eq.(\ref{eqn:Sal}) one arrives at the reduced
Salpeter-equation, which then has the from of a Schr\"odinger equation
with relativistic kinetic energy and relativistic corrections to the
potential (contained in $\Lambda^{\pm}$). This can be considered the
starting point of virtually all ``relativized'' constituent quark
models.

Pioneering work in this spirit was performed already almost two
decades ago by the group around Nathan Isgur both for mesons,
see~\cite{god85}, and later on for baryons, see~\cite{cap00} and
references therein. Here it was assumed that the quark interactions in
hadrons can effectively be described by a linear confinement potential
and spin-dependent parts of one gluon exchange; relativistic effects
in the interactions were accounted for by parametrizations. This also
holds for the description of annihilation contributions to
pseudoscalar mixings. The scope of the calculation e.g. for mesons is
a unified description of all resonances, both with light and with
heavy flavours, and also includes a calculation of a multitude of
electroweak and strong decay observables, which in spite of the more
than a dozen model parameters can still be considered as rather
efficient.

\begin{figure}[htb]
  \includegraphics[height=0.4\textheight]{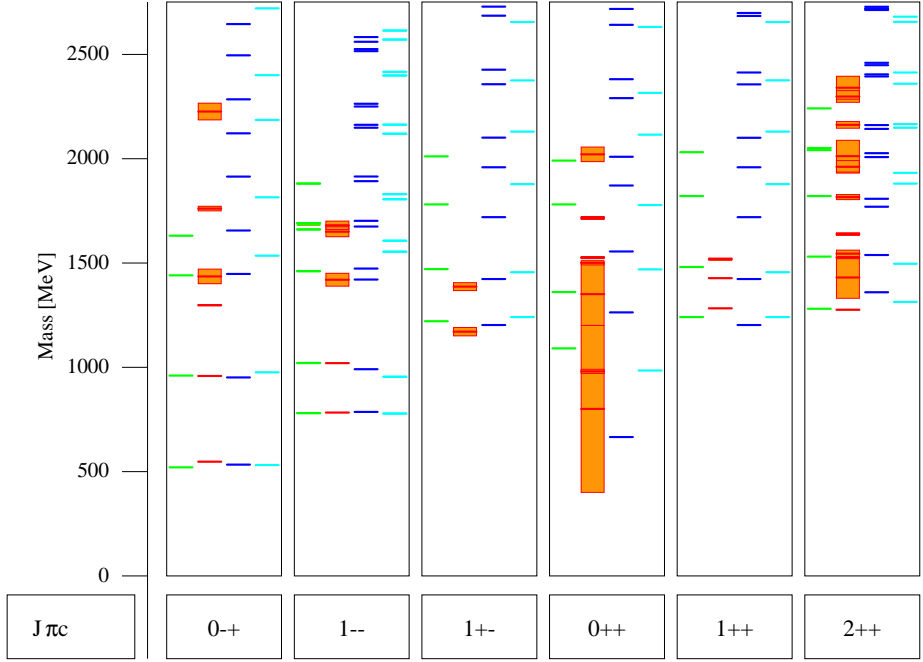}
  \caption{Spectrum of $S$- and $P$ wave isoscalar mesons. From left
  to right each column (of fixed spin $j$, parity $\pi$ and charge
  parity $c$) displays the results from the Godfrey-Isgur
  'relativized' calculation~\protect{\cite{god85}}, 
  the experimental resonance position with a box
  indicating the error, and two versions of the relativistic
  calculation on the basis of the Salpeter equation with
  instanton-induced forces~\protect{\cite{kol00}}.}
  \label{metsch:fig:one} 
\end{figure}

\begin{figure}[htb]
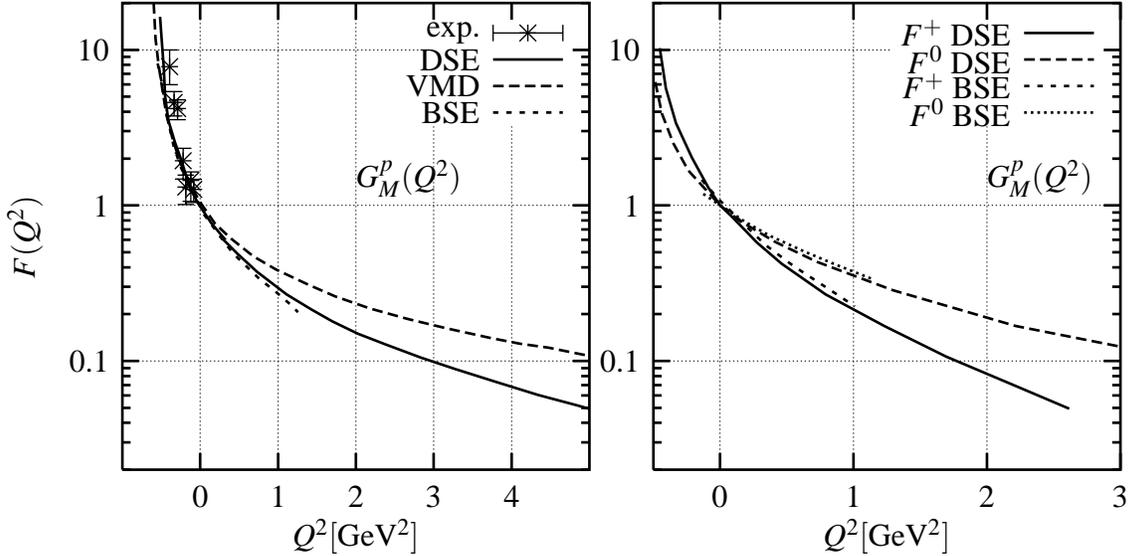

  \hspace*{4em} 
  \input{omegapigamma.pstex}
  \hspace*{-4em} 
  \input{KstarKgamma.pstex} 
  \caption{Left: comparison of $\omega-\pi-\gamma$-transition form factor
	calculated in the Dyson-Schwinger approach
  (DSE) of
  \protect{\cite{mar03}} with experimental data in the time-like
  region, the results from the instantaneous Bethe-Salpeter approach
  (BSE) and simple $\omega$-vector meson dominance; Right: predictions for
  the charged ($F^+$) and neutral ($F^0$) $K^*-K-\gamma$-transition 
  form factors}
  \label{metsch:fig:two} 
\end{figure}

On the other hand one can also take the full Salpeter equation as a
starting point for constituent quark model calculations: here the
instantaneous interaction kernel consists of the Fourier transform of
a string-like linearly rising confinement potential with an
appropriate Dirac structure which avoids large spin-orbit splittings,
supplemented by a spin-flavour dependent interaction motivated by
instanton effects, see \cite{kol00}.  The latter has the decisive
advantageous property to incorporate the $U_A(1)$ anomaly
quantitatively and thus to account immediately for the splitting and
mixing of (pseudo)scalar mesons. The total number of parameters in
this approach amount to seven. As an example a comparison of the
isoscalar mass spectrum for two versions of the confinement potential
(Model $\mathcal{A}$ and Model $\mathcal{B}$ employing confinement
Dirac structures
$(\frac{1}{2}\left(\Id\otimes\Id-\gamma_0\otimes\gamma_0\right)$ 
and
$\frac{1}{2}\left(\Id\otimes\Id-\gamma_5\otimes\gamma_5-\gamma^\mu
\otimes \gamma_\mu\right)$, 
respectively) with experimental data and the results from the
calculation of Godfrey and Isgur is given in Fig.~\ref{metsch:fig:one}.    
Apart from the scalar sector the results are rather similar. While the
'relativized' quark model calculation resort to the rather \textit{ad
hoc} 'mock-meson' method, in the field theoretical approaches based on
the Bethe-Salpeter equation the calculation of decay amplitudes in the
Mandelstam-formalism is straightforward and parameter free, albeit
numerically tedious.  A comparison of the results for pseudoscalar
decay constants is given in Table~\ref{metsch:tab:one}, 
\begin{table}[htb]
\begin{tabular}{cccccc}
\hline
  & \tablehead{1}{c}{b}{Model $\mathcal{A}$}
  & \tablehead{1}{c}{b}{Model $\mathcal{B}$}
  & \tablehead{1}{c}{b}{DSE}
  & \tablehead{1}{c}{b}{Exp.}
  & \tablehead{1}{c}{b}{OGE}   \\

\hline
$f_\pi$ & 212 & 219 & 132 & $130.7 \pm 0.46$ & 184\\
$f_K$   & 248 & 238 & 154 & $159.8 \pm 1.84$ & 235 \\
\hline
\end{tabular}
\caption{Pseudoscalar decay constants in [MeV]}
\label{metsch:tab:one}
\end{table}
for some radiative transitions in~\ref{metsch:tab:two} 
\begin{table}[htb]
\begin{tabular}{cccccc}
\hline
    \tablehead{1}{c}{b}{Decay}
  & \tablehead{1}{c}{b}{Model $\mathcal{A}$}
  & \tablehead{1}{c}{b}{Model $\mathcal{B}$}
  & \tablehead{1}{c}{b}{DSE}
  & \tablehead{1}{c}{b}{Exp.}
  & \tablehead{1}{c}{b}{OGE}   \\

\hline
$\rho^\pm\to\pi^\pm\gamma$ & 35 & 21 & 53 & $67 \pm 9$ & 67\tablenote{fitted}\\
$\rho^0  \to\pi^0  \gamma$ & 35 & 21 &    &$117\pm 30$ & 67\\
$\rho^0  \to\eta   \gamma$ & 50 & 40 &    & $57\pm 11$ & 51\\
$\omega  \to\pi^0  \gamma$ &315 &185 &479 &$717\pm 42$ &642\\
$\omega  \to\eta   \gamma$ &5.5 &4.4 &    &$5.5\pm 0.8$&5.4\\
$K^{*\pm}\to K^\pm \gamma$ & 48 & 29 & 90 & $50\pm 5  $& 67\\
$K^{*0}  \to K^0   \gamma$ &102 & 70 &130 &$117\pm 10 $&118\\
$\eta'   \to\rho^0 \gamma$ & 87 & 28 &    & $60\pm 5 $ &135\\
$\eta'   \to\omega \gamma$ &9.7 &3.1 &    &$6.1\pm 0.8$&13.4\\
$\phi    \to\eta   \gamma$ & 58 & 35 &    &$ 58\pm 2$  &66\\
$\phi    \to\eta'  \gamma$ &0.01&0.08&    &$0.30\pm0.16$&0.26\\
\hline
\end{tabular}
\caption{Decay widths in [keV] of radiative meson transitions}
\label{metsch:tab:two}
\end{table}
and of the $\omega \to \pi\gamma$ and $K^* \to K \gamma$ transition
form factors in Fig.~\ref{metsch:fig:two}. 
The Dyson-Schwinger approach leads to an excellent description of some 
observables, but at the time is unfortunately limited to calculations 
on properties of the lowest pseudoscalar and vector mesons mainly. 
This restriction does not applies to the instantaneous
Bethe-Salpeter approach, which simultaneously describes the whole mass
spectrum and without introducing new parameters does fairly well also
for observables at higher momentum transfer, see~\cite{kol00}. 
Such observables were not calculated in the relativized quark model, 
nevertheless the \textit{ad hoc} ``mock-meson method'' gives a 
remarkable description of a multitude of other experimental data.

\section{Baryons}
\begin{figure}[htb]
  \includegraphics[height=0.3\textheight,width=0.5\textwidth]{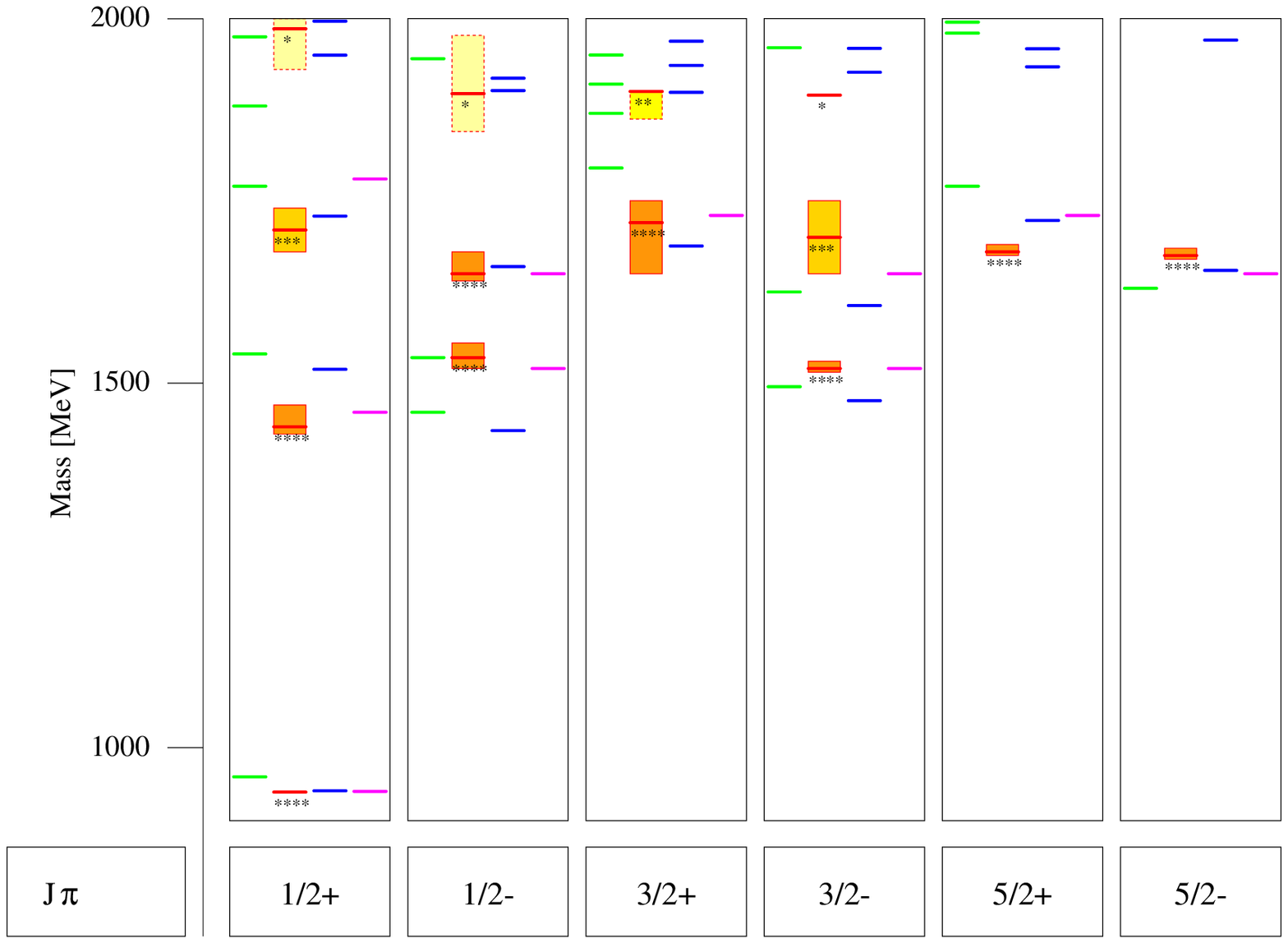}
  \includegraphics[height=0.3\textheight,width=0.5\textwidth,clip]{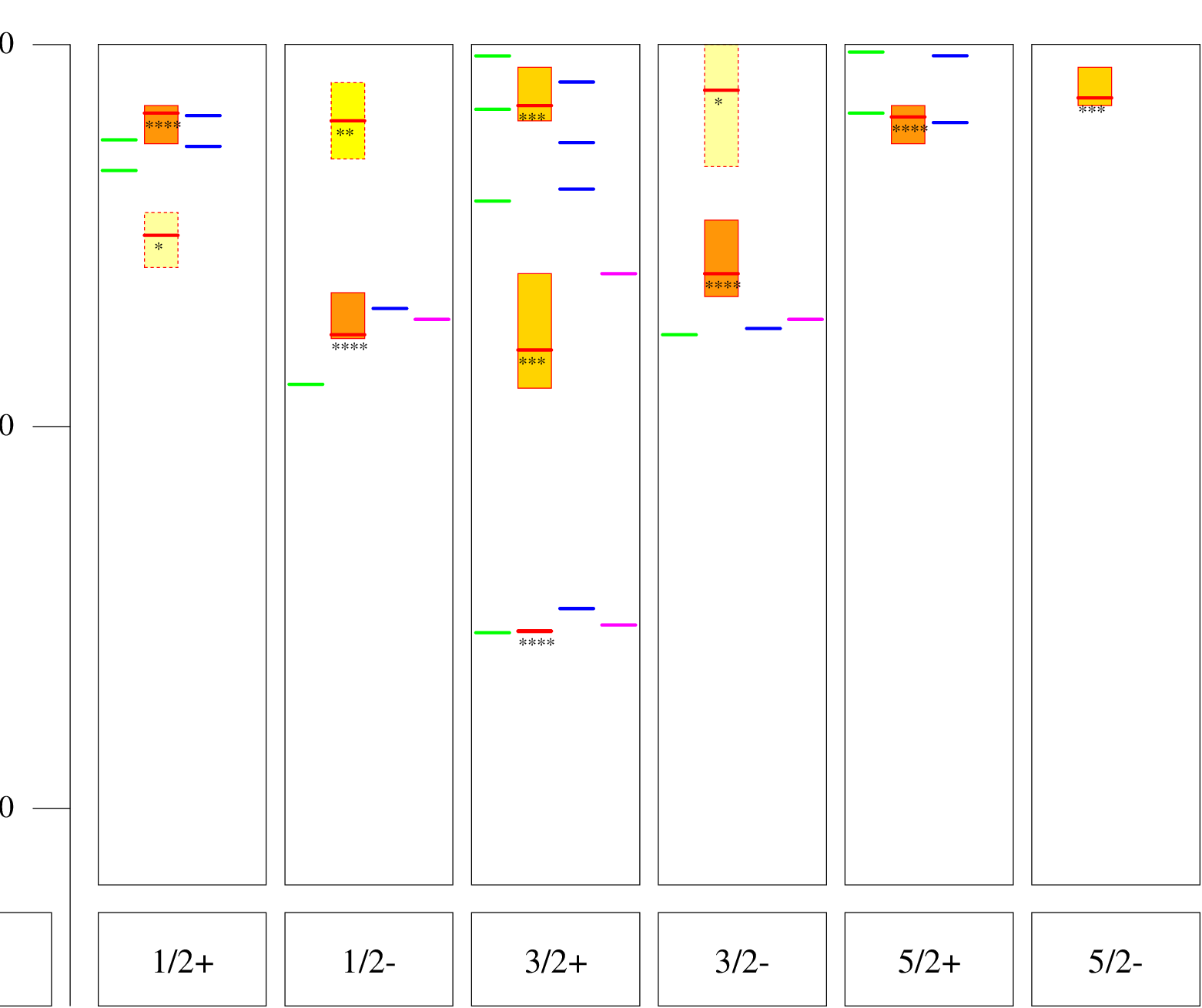}
  \caption{Low lying $N$- (left) and $\Delta$-resonances
  (right). In each column from left to right the calculated result
  from the relativized quark model with a OGE-based quark
  interaction~\protect{\cite{cap00}},
  the experimental data, the result from the instantaneous
  Bethe-Salpeter Equation with instanton-induced interactions and the
  results from a quark model calculation with Goldstone boson
  exchange~\protect{\cite{ple03}} is displayed.  
}
\label{metsch:fig:three}
\end{figure}

Although some pilot studies on (ground states of) baryons as
$q^3$-systems have been done in the Dyson-Schwinger approach within a
diquark-quark picture~\cite{mar03}, the majority of constituent quark
models of baryons still rely on the non-relativistic treatment with
(some) relativistic corrections. If one insists on a description of
the whole mass spectrum implementing relativistic covariance both in
the quark dynamics and in the calculation of currents needed for decay
observables, again, as for mesons, the instantaneous Bethe-Salpeter
equation seems to be an appropriate starting point. Baryons are thus
described by the homogeneous Bethe-Salpeter equation with
three-particle and two-particle instantaneous interaction kernels,
which implement confinement and a spin-dependent interaction to
account for the major mass splittings. Again the assumption of
effective constituent quark propagators of the free form together with
an approximate treatment of the two-body interactions~\cite{loe01}
allows to formulate the dynamics in terms of Salpeter amplitudes
(e.g. in the rest frame of the baryons):
$\Phi_M(\vec p_\xi,\vec p_\eta :=
\int\!\!\frac{dp_\xi^0}{2\pi}\frac{dp_\eta^0}{2\pi}\chi_M(p_\xi,p_\eta)  
$
which then fulfills the Salpeter equation:
\begin{eqnarray}
\lefteqn{(\mathcal{H}\Phi_M)({\vec p_\xi},{\vec p_\eta})
=\sum_{i=1}^3 {H_i}\;\Phi_M({\vec p_\xi},{\vec p_\eta})}&&\nonumber\\
&+&
\left(
{\Lambda_1^+}\otimes{\Lambda_2^+}\otimes{\Lambda_3^+}
+
{\Lambda_1^-}\otimes{\Lambda_2^-}\otimes{\Lambda_3^-}\right)\nonumber\\
&&
\;\;\;\gamma^0\otimes\gamma^0\otimes\gamma^0
\int\frac{d^3\!p_\xi'}{(2\pi)^3}\;\frac{d^3\!p_\eta'}{(2\pi)^3}\;
V^{(3)}({\vec p_\xi},{\vec p_\eta},{\vec p_\xi'},{\vec p_\eta'})\;
\Phi_M({\vec p_\xi'},{\vec p_\eta'})\nonumber\\
&+&\left({\Lambda_1^+}\otimes{\Lambda_2^+}\otimes{\Lambda_3^+}
-{\Lambda_1^-}\otimes{\Lambda_2^-}\otimes{\Lambda_3^-}\right)\nonumber\\
& &
\;\;\;\gamma^0\otimes\gamma^0\otimes\Id\;
\int\frac{d^3\!p_\xi'}{(2\pi)^3}\;
\left[V^{(2)}({\vec p_\xi},{\vec p_\xi'})\otimes \Id\right]
\;\Phi_M({\vec p_\xi'},{\vec p_\eta})\nonumber\\
&& +\;\; \textrm{cycl. perm. (123)}\,.
\label{metsch:eq:two} 
\end{eqnarray}
Again, if one would drop all terms involving the negative energy
projectors $\Lambda^-$ one arrives at a Schr\"odinger-type equation
with relativistic corrections. Although the full Salpeter hamiltonian
(\ref{metsch:eq:two}) is not positive definite with respect to the
scalar product of the Salpeter amplitudes and thus positive and
negative energy solutions occur, the negative energy solutions can
(via the CPT-transformation) be mapped to positive energy solutions of
opposite parity and consequently this approach leads to the same number
of states as the non-relativistic quark model. Again, adopting a
linear three-body confinement potential with a suitable spin
dependence avoiding large spin-orbit splittings and the
instanton-induced interaction to account for the major spin-dependent
splittings with only seven parameters an excellent description has
been obtained for all light-flavoured baryons,
see~\cite{loe01a,loe01b}, including selective parity doubling and the
Regge-trajectories up to the highest measured masses and total angular
momenta. In~\ref{metsch:fig:three}
the results for $N$- and $\Delta$-resonances of low-mass and low
angular momenta is compared to experimental data as well as to the
results from the relativized constituent quark model using parts of
the OGE as a residual interaction, see~\cite{cap00} and the results
from a constituent quark model developed by the Graz-group, which
employs (flavour dependent) modified Yukawa-type potentials based on
Goldstone-Boson-Exchange, see e.g. ~\cite{ple03}.  The latter
treatment has the obviously satisfactory feature to be able to
reproduce the first excited states of positive parity below the
negative parity states, whereas the other treatments do yield a low
lying Roper-like resonance but slightly above the lowest negative
parity states. All calculations can not account for some negative
parity $\Delta$-resonances at approximately 1.9 GeV, see also the
contribution of Ch. Weinheimer to this conference.

\begin{figure}[htb]
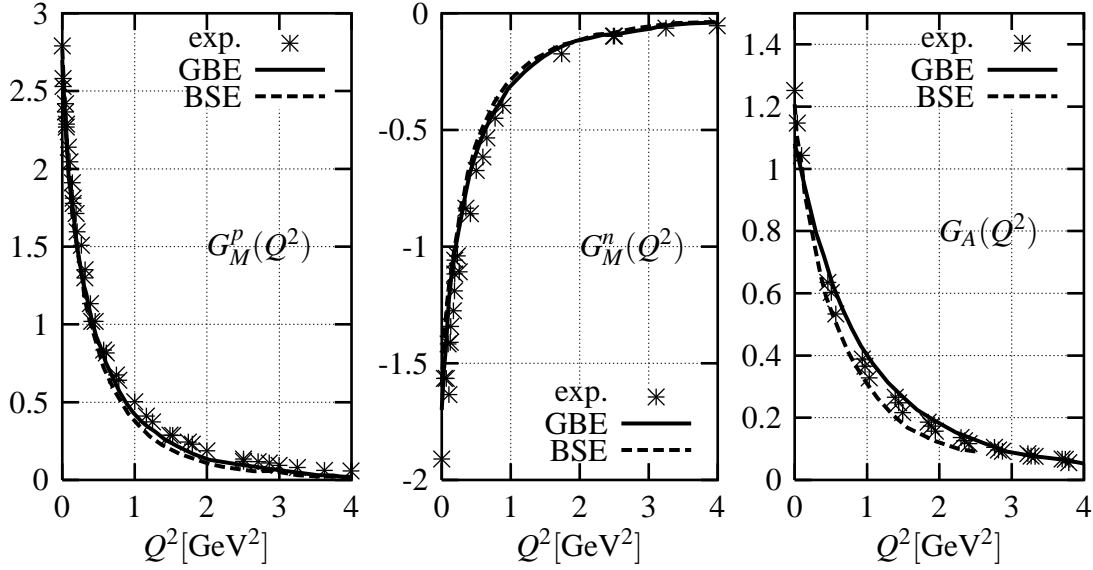

  \hspace*{-1.0em} 
  \input{GMp.pstex}
  \hspace*{-1.0em} 
  \input{GMn.pstex}
  \hspace*{-1.0em} 
  \input{GA.pstex}
  \caption{Comparison of the
  form factors calculated in the point-from approach in the 
  constituent quark model with Goldstone
  Boson Exchange (GBE) of \protect{\cite{ple03}} and in the Mandelstam
  formalism on the basis of the Bethe-Salpeter Equation (BSE)
  \protect{\cite{loe02}}  
  with experimental data; Left: Magnetic form
  factor of the proton; middle: Magnetic form factor of the neutron
  and right: axial form factor (adapted from \protect{\cite{ple03}})}
  \label{metsch:fig:four} 
\end{figure}

As for the mesons, in the Bethe-Salpeter approach electroweak currents
can be calculated covariantly and (in lowest order) parameter free
within the Mandelstam formalism~\cite{loe02}.  The results for the
magnetic moments of octet and decuplet baryons are given in
Table~\ref{metsch:tab:three} together with experimental data and the
results which the Graz-group obtained employing the point-form of
Dirac's relativistic quantum mechanics.
\begin{table}[htb]
\begin{tabular}{lrrrlrrr}
\hline
    \tablehead{1}{c}{b}{Baryon}
  & \tablehead{1}{c}{b}{BSE}
  & \tablehead{1}{c}{b}{Exp.}
  & \tablehead{1}{c}{b}{GBE}
  & \tablehead{1}{c}{b}{Baryon}
  & \tablehead{1}{c}{b}{BSE}
  & \tablehead{1}{c}{b}{Exp.}
  & \tablehead{1}{c}{b}{GBE}   \\
\hline
$p$ 	& 2.77	& 2.793	& 2.70 & $\Xi^0$	& -1.33 & -1.250 & -1.27 \\
$n$	&-1.71  &-1.913 & -1.70& $\Xi^-$  	& -0.56 & -0.6507& -0.67 \\
$\Lambda$&-0.61 &-0.613 & -0.59& $\Delta^+$	& 2.07  & 
\multicolumn{1}{c}{$2.7\pm 1.5 \pm 1.3$} 
& 2.08 \\
$\Sigma^+$&2.51 &2.458  & 2.34 & $\Delta^{++}$     & 4.14  & 3.7-7.5&
    4.17 \\
$\Sigma^-$&-1.02&-1.160 & -0.94& $\Omega^-$     & -1.66 & -2.0200 &
    -1.59\\
$\Sigma^0$&0.75 & -- & 0.70 &                   &       &         & \\
\hline
\end{tabular}
\caption{Magnetic moments in $\mu_N$ of octet and decuplet baryons}
\label{metsch:tab:three}
\end{table}
Although the calculational frameworks and the quark dynamics differ
substantially in both approaches the results are remarkably similar
and stress the importance of a relativistically covariant calculation
of electromagnetic currents. This holds \textit{a forteriori} for the
calculation of electromagnetic (transition) form factors, see
\textit{e.g.} the comparison in Fig.~\ref{metsch:fig:four}.
For more results on electroweak transition form factors we refer
to~\cite{loe02}. Some new, representative results for semi-leptonic
decays, calculated from the weak baryonic currents in the Mandelstam
formalism, are listed in Table~\ref{metsch:tab:four}.
\begin{table}[htb]
\newcommand{\m}{\hphantom{$-$}}
\newcommand{\cc}[1]{\multicolumn{1}{c}{#1}}
\begin{tabular}{lclllll}
\hline
  &&
  & \tablehead{2}{c}{b}{$\Gamma$ $[10^{6}\textrm{s}^{-1}]$}
  & \tablehead{2}{c}{b}{$g_A/g_V$}\\
    \tablehead{3}{c}{b}{Decay}
  & \tablehead{1}{c}{b}{Exp.}
  & \tablehead{1}{c}{b}{Calc.}
  & \tablehead{1}{c}{b}{Exp.}
  & \tablehead{1}{c}{b}{Calc.} \\
\hline
$n$ 	   &$\to$& $p\,e^-\bar\nu_e$   	        &                       &           &\m{$1.2670 \pm 0.0035$}    &\m{$1.21$}\\
$\Lambda$  &$\to$& $ p\,e^-\bar\nu_e$		&\cc{$3.16 \pm 0.06$}   &\cc{$3.10$}&{$-0.718 \pm 0.015$}       &{$-0.82$}\\
$\Sigma^+$ &$\to$& $\Lambda \,e^+\nu_e$		&\cc{$0.25 \pm 0.06$}	&\cc{$0.20$}& & \\
$\Sigma^-$ &$\to$& $\Lambda \,e^-\bar\nu_e$	&\cc{$0.38 \pm 0.02$}  	&\cc{$0.34$}& & \\
$\Sigma^-$ &$\to$& $ n\,e^-\bar\nu_e$		&\cc{$6.9 \pm 0.2$}    	&\cc{$4.91$}&\m{$0.340 \pm 0.017$}      &\m{$0.25$}	\\
$\Xi^0$    &$\to$& $\Sigma^+\,e^-\bar\nu_e$	&\cc{$0.93\pm 0.14$}   	&\cc{$0.91$}&\m{$1.32 {+0.21 \atop -0.17} \pm 0.05$} &\m{$1.38$}\\
$\Xi^-$    &$\to$& $\Sigma^0\,e^-\bar\nu_e$	&\cc{$0.5 \pm 0.1$}    	&\cc{$0.51$}& &	\\
$\Xi^-$    &$\to$& $\Lambda \,e^-\bar\nu_e$	&\cc{$3.3 \pm 0.2$}    	&\cc{$2.30$}&{$-0.25 \pm 0.05$}         &{$-0.27$}\\
$\Omega^-$ &$\to$& $ \Xi^0\,e^-\bar\nu_{e}$	&\cc{$68 \pm 34$}      	&\cc{$46$}  & &	\\
$\Lambda$  &$\to$& $ p\,\mu^-\bar\nu_{\mu}$	&\cc{$0.60 \pm 0.13$}  	&\cc{$0.47$}& &	\\
$\Sigma^-$ &$\to$& $ n\,\mu^-\bar\nu_{\mu}$	&\cc{$3.04 \pm 0.27$}  	&\cc{$1.60$}& &	\\
$\Xi^-$    &$\to$& $\Lambda\,\mu^-\bar\nu_{\mu}$&\cc{$2.1 \pm 1.3$}   	&\cc{$1.04$}& &	\\
\hline
\end{tabular}
\caption{Decay rates and axial vector couplings of semi-leptonic decays of baryons.}
\label{metsch:tab:four}
\end{table}
Electroweak currents, provided that they are calculated in a
relativistically covariant framework, can thus be satisfactory
calculated in lowest order.

This is no longer holds \textit{a priori} for the calculation of
strong two-body decays, where channel couplings and mixing can be
important, and in principle resonances could be even generated
dynamically through such effects. Nevertheless it seems interesting to
investigate to what extent a lowest order calculation, without any
introduction of new parameters can describe some experimental
features.

\begin{minipage}[c]{0.25\textwidth}
\begin{center}
\includegraphics[width=1.0\textwidth]{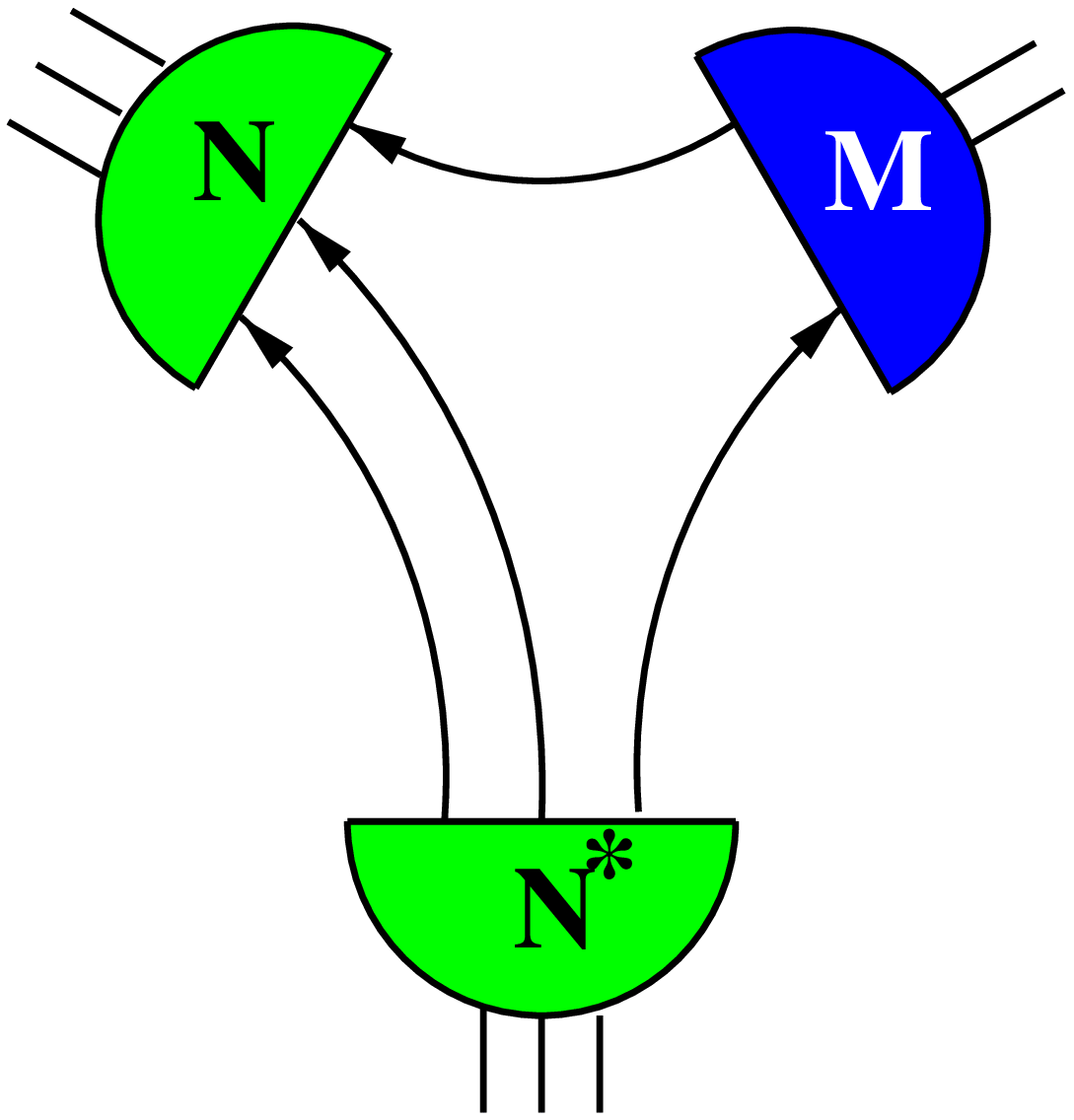}\\
\end{center}
\end{minipage}
\hspace{\fill}
\begin{minipage}[c]{0.65\textwidth}

In the framework of the Mandelstam formalism the amplitude for the
strong mesonic decay of excited baryons can be obtained in lowest
order by evaluating the simple quark loop diagram displayed on the
left, which involves the vertex functions (amputated
Bethe-Salpeter-amplitudes) of the participating meson, obtained from
the calculation on mesons~\cite{kol00}, and of the initial and final
baryon.
\end{minipage}

\begin{figure}[htb]
\includegraphics[angle=270,width=1.0\textwidth]{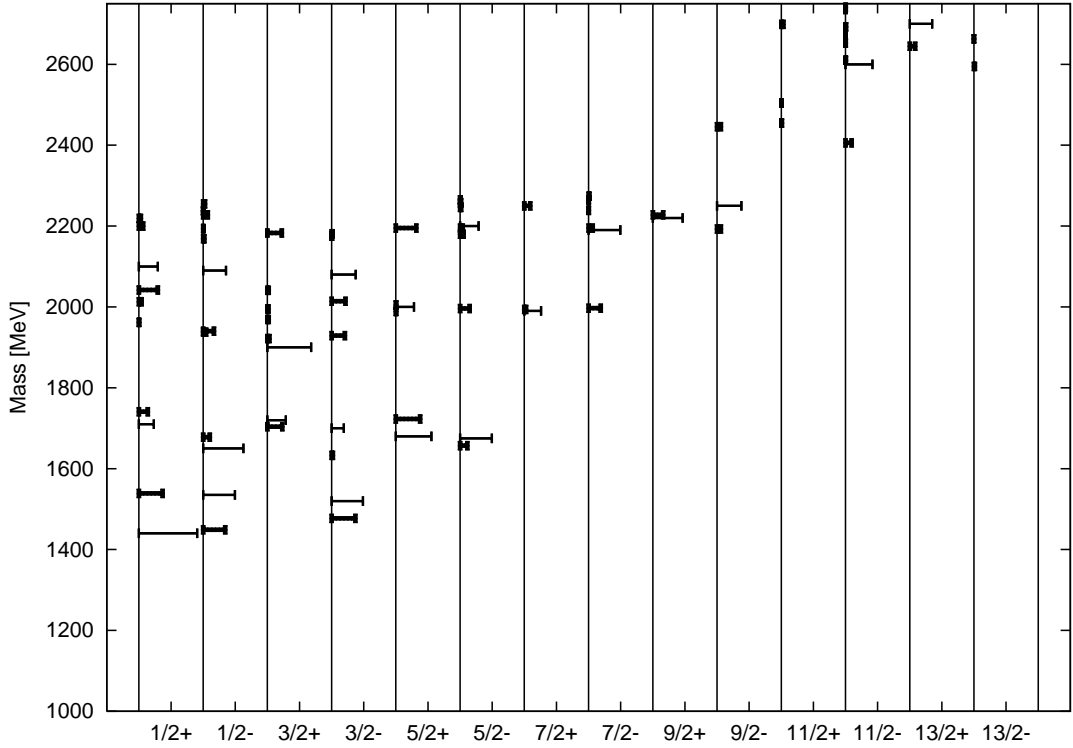}
\caption{
Decay amplitudes (proportional to the square root of the partial decay
width) of strong $N^* \to N\pi$ decays. In each column (\textit{i.e.}
for each spin and parity $J\pi$) the experimental value (thin
horizontal bars at the experimental resonance position) is compared to
the calculated value (thick horizontal bars at the calculated
resonance position).
}
\label{metsch:fig:six}
\end{figure}

Although in general the calculated partial widths are too small to
account for the experimental values quantitatively, appreciable decay
widths are found only for the well established resonances, the
predicted values for higher lying resonances being in general smaller
by at least an order of magnitude, see also Fig.~\ref{metsch:fig:six},
thus explaining why these have not been observed so far in elastic
pion-nucleon scattering. This observation is in accordance with
previous findings cited in~\cite{cap94,cap00}\,.  In
Table~\ref{metsch:tab:five} the calculated partial decay widths of
some selected low lying $N$- and $\Delta$-resonances are compared to
experimental data and recent results from the Graz group with a
relativistic elementary meson emission~\cite{ple03} (with no extra
parameters) as well as the results from the relativized quark model
invoking the ${}^3P_0$-model~\cite{cap94}.
\begin{table}[htb]
\begin{tabular}{lcccc@{\hspace{-0.2em}}clccc@{\hspace{-0.2em}}c}
  \tablehead{1}{c}{b}{Decay}
  & \tablehead{1}{c}{b}{BSE}
  & \tablehead{1}{c}{b}{GBE}
  & \tablehead{1}{c}{b}{${}^3P_0$}
  & \tablehead{2}{c}{b}{Exp.}
  & \tablehead{1}{c}{b}{Decay}
  & \tablehead{1}{c}{b}{BSE}
  & \tablehead{1}{c}{b}{${}^3P_0$}
  & \tablehead{2}{c}{b}{Exp.}\\
\hline
$S_{11}(1535)\to N\pi$ & {$33$}  & {$93$}  & {$216$}   & {$(68\pm 15)$} & {${+45\atop -23}$} &
$\to \Delta\pi$        & {$1$}   & {$2$}     & {$<2$}         &                  \\
$S_{11}(1650)\to N\pi$ & {$3$}   & {$29$}  & {$149$}   & {$(109\pm 26)$}& {${+29\atop -4} $} &
$\to \Delta\pi$        & {$5$}   & {$13$}    & {$(6\pm 5)$}   & {${+2\atop 0}$}    \\
$D_{13}(1520)\to N\pi$ & {$38$}  & {$17$}  & {$74$}    & {$(66\pm 6)$}  & {${+8\atop -5}$}   &
$\to \Delta\pi$        & {$35$}  & {$35$}    & {$(24\pm 6)$}  & {${+3\atop -2}$}   \\
$D_{13}(1700)\to N\pi$ & {$0.1$} & {$1$}  & {$34$}    & {$(10\pm 5)$}  & {${+5\atop -5}$}   &
$\to \Delta\pi$        & {$88$}  & {$778$}   & {seen}         &                  \\
$D_{15}(1675)\to N\pi$ & {$4$}   & {$6$}  & {$28$}    & {$(68\pm 7)$}  & {${+14\atop -5}$}  &
$\to \Delta\pi$        & {$30$}  & {$32$}    & {$(83\pm 7)$}  & {${+17\atop -6}$}  \\
\hline
$P_{11}(1440)\to N\pi$ & {$38$}  & {$30$}  & {$412$}   & {$(228\pm 18)$}& {${+65\atop -65}$} &
$\to \Delta\pi$	       & {$35$}  & {$11$}    & {$(88\pm 18)$} & {${+25\atop -25}$} \\
\hline
$P_{33}(1232)\to N\pi$ & {$62$}  & {$34$}  & {$108$}   & {$(119\pm 0)$} & {${+5\atop -5}$}   &
                       &       &         &              &                  \\ 
\hline
$S_{31}(1620)\to N\pi$ & {$4$}   & {$10$}  & {$26$}    & {$(38\pm 7)$}  & {${+8\atop -8}$}   &
$\to \Delta\pi$        & {$72$}  & {$18$}    & {$(68\pm 23)$} & {${+14\atop -14}$} \\
$D_{33}(1700)\to N\pi$ & {$2$}   & {$3$}  & {$24$}    & {$(45\pm 15)$} & {${+15\atop -15}$} &
$\to \Delta\pi$        & {$52$}  & {$262$}   & {$(135\pm 45)$}& {${+45\atop -45}$} \\
\hline			                                               
\end{tabular}
\caption{Partial decay widths in MeV 
of some strong two-body decays of $N$-
and $\Delta$-resonances.}
\label{metsch:tab:five}
\end{table}

\section{Conclusion}

In conclusion we think that we have demonstrated, that constituent quark models provide a very useful tool
in understanding hadron properties in a unified way: This not only
involves a description of the mere mass spectrum, but also numerous
decay amplitudes and electroweak (transition) form factors. In
particular the field theoretical approaches which rely on the
description of bound states of quarks through coupled Bethe-Salpeter/
Dyson-Schwinger equations have provided very interesting results,
unfortunately so far only for the ground states and some low-lying
excited states. In this respect the approach based on the
instantaneous Bethe-Salpeter equation, using free-form fermion
propagators with constituent masses, implementing confinement by a
string-like linearly rising potential and with instanton-induced
interactions to explain the spin-dependent mass splittings seems 
to be a very efficient compromise combining the advantages of a 
relativistically covariant
field theoretical treatment with the successful concepts 
of the (non-relativistic) constituent quark model. 
Adopting the point form of Dirac's
Relativistic Quantum Mechanics does improve the description of
observables within the latter category drastically, and supports the
main findings of our treatment that a relativistic treatment of decay
amplitudes, especially for processes at higher momentum transfers is
absolutely imperative.    

\begin{theacknowledgments}
I like to acknowledge the longstanding fruitful collaboration with Herbert
  Petry and the contributions by Matthias Koll, Ulrich L\"oring, Dirk Merten, 
Christian Haupt and Sascha
  Migura who did most of the calculations.
\end{theacknowledgments}

\bibliographystyle{aipproc}

\end{document}